\documentstyle[10pt,epsf,subfigure]{elsart} 
 
\begin{document} 
 
\begin{frontmatter} 
\title{Weak binding limit and non zero angular momentum states 
in Light-Front Dynamics} 
\author{M. Mangin-Brinet$^a$, J. Carbonell$^a$ and V.A. Karmanov$^b$} 
 
\address{$^a$ Institut des Sciences Nucl\'{e}aires, 
        53 Av. des Martyrs, 38026 Grenoble, France} 
\address{$^b$ Lebedev Physical Institute, Leninsky prospekt 53, 
117924 Moscow, Russia} 
 
\begin{abstract} 
We show some results concerning the weak binding limit for  
$J=0$ states -- which turn out to strongly differ from the non  
relativistic case -- together with the construction of non zero angular  
momentum states. The calculation of such states in the  
Light-Front Dynamics (LFD) framework has some peculiarities  
which are absent in other approaches. They are related to the fact that  
the rotation generators contain interaction. We present here the  
construction of non zero angular momentum states in LFD and show how it  
leads to a restoration of rotational invariance. For this purpose, the  
use of Light-Front Dynamics in its explicitly covariant formulation  
\cite{CDMK_98} is of crucial importance since the dependence of the  
wave function on the light-front plane is explicitly parametrized.   
\end{abstract} 
\end{frontmatter} 
 
\vspace{-0.0cm} 
In the explicitly covariant version of LFD, the state vector is defined on a plane 
whose equation is written in the form: $\omega\cdot x=\sigma$, where 
$\omega$ is a  
light-like four-vector. The standard approach of LFD \cite{BPP_98} is recovered for 
$\omega=(1,0,0,-1)$. 
The dynamical equations for the state vector  
describing a physical system are obtained by imposing that it obeys the  
Poincar\'e group transformation laws and that it belongs to an  
irreducible representation. 
In addition, we are interested in systems with definite  
four-momentum $p^{\mu}$ and angular momentum projection $\lambda$, that  
is $\Psi$ satisfies: 
\vspace{-0.1cm} 
\begin{eqnarray}  
&&\hat{P}^2 \Psi =M^2 \Psi, \qquad \qquad \qquad \! 
\hat{P}^{\mu}\Psi = p^{\mu}\Psi,  
\label{P1}\\  
&&\hat{S}^2 \Psi = -M^2 J(J+1) \Psi, \quad \; 
\hat{S}_{3} \Psi = M \lambda\Psi. 
\label{S1}   
\end{eqnarray}  
\vspace{-0.1cm} 
$\hat{P}^{\mu}$ is the four-momentum operator and $\hat{S}^{\mu}$ 
the Pauli-Lubansky one, given by 
$$ 
\vspace{-0.1cm} 
\hat{P}^{\mu} = \int T^{\mu\nu}(x) \,\delta(\omega\cdot x-\sigma)\, 
\omega_{\nu}\,d^4x,  \qquad  
\hat{S}_{\mu}={1\over 2} \varepsilon_{\mu\nu\rho\sigma} \hat{P}^{\nu} 
\hat{J}^{\rho\sigma}  
\vspace{-0.0cm} 
$$ 
where $T^{\mu\nu}$ is the energy momentum tensor and $M$ the total mass  
of the system.   
 
We consider a model consisting of two interacting scalar fields  
$\varphi$ and $\chi$ with masses $m$ and $\mu$ respectively  
\cite{MC_00}. The interaction Hamiltonian is  
${\cal{H}}=-g\varphi^2\chi$. The wave functions are the Fock components  
of the state vector and the kernels are restricted to the usual  
ladder approximation. We are looking for solutions written on a form  
which directly satisfies the four-momentum eigenvalue equation in  
(\ref{P1}). The mass eigenvalue equation in (\ref{P1}) results into a  
three-dimensional integral equation for the two-body Fock component  
$\psi(\vec{k},\hat{n})$: 
\vspace{-0.2cm} 
\begin{eqnarray*} 
[4(\vec{k}^2 +m^2)-M^2] \psi(\vec{k},\hat{n}) 
=-{m^2 \over 2 \pi^3} \int {d^3k' \over 
\varepsilon_{k'}  
}V(\vec{k},\vec{k}\,',\hat{n},M)\psi(\vec{k}\,',\hat{n})  
\end{eqnarray*} 
$\vec{k}$ is the momentum of one particle in the system of reference where  
$\vec{k}_1+\vec{k}_2=\vec{0}$, $\hat{n}$ is the unit vector in the  
direction of $\vec{\omega}$ and $\varepsilon_{k}=\sqrt{\vec{k}^2+m^2}$.  
For $J=0$ state, $\psi$ is a scalar function depending on the scalars  
$k$ and $\hat{k}\cdot\hat{n}$.  The kernel in the r.h.-side can be  
integrated analytically over the azimuthal angle and the equation is  
reduced to a two-dimensional one. 
\begin{figure}[htbp] 
\vspace{-0.3cm} 
\begin{center} 
{\mbox{\epsfxsize=6.6cm\epsfysize=7.2cm 
\subfigure[ ]{\epsffile{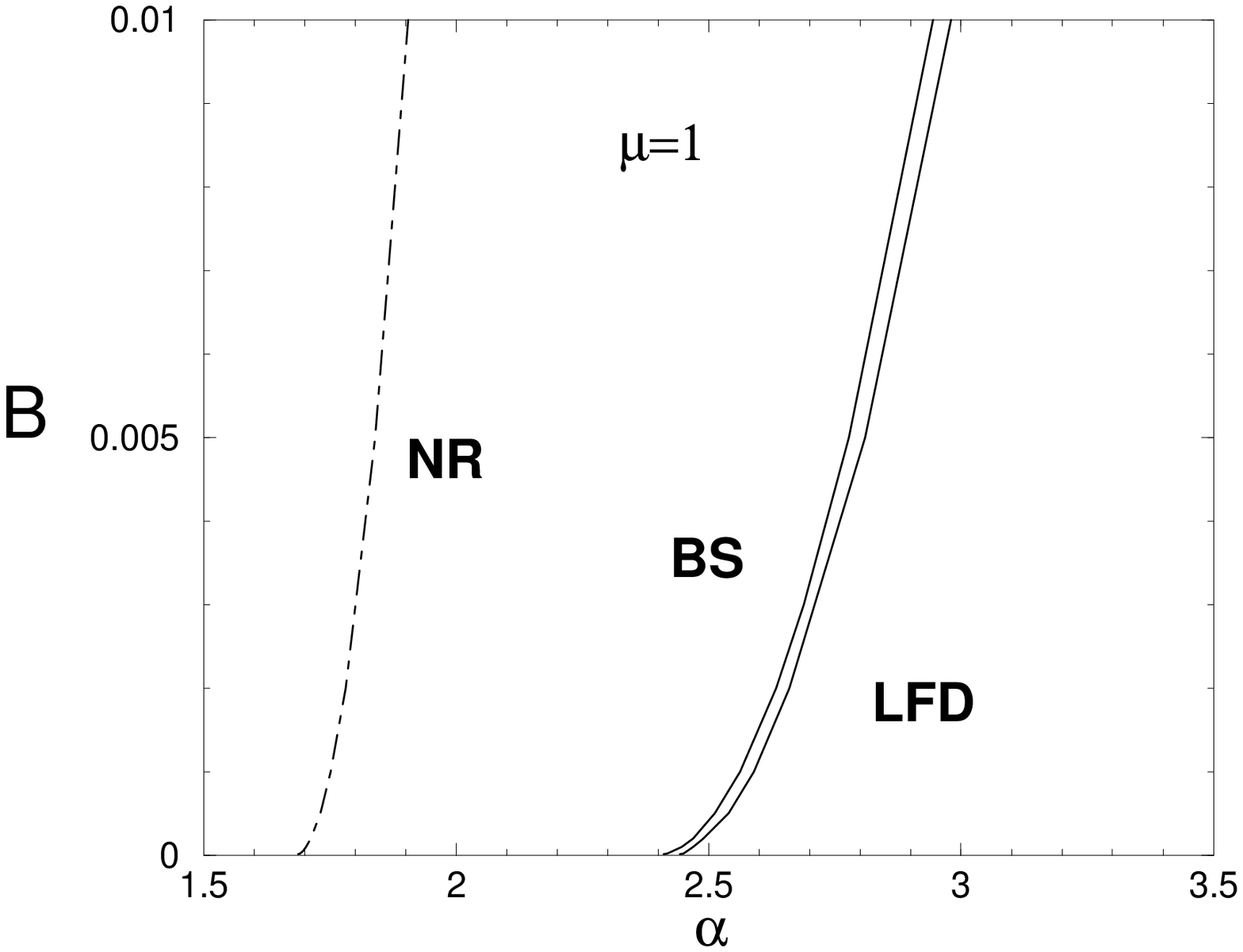}}}} 
{\mbox{\epsfxsize=6.6cm\epsfysize=8.1cm 
\subfigure[ ]{\epsffile{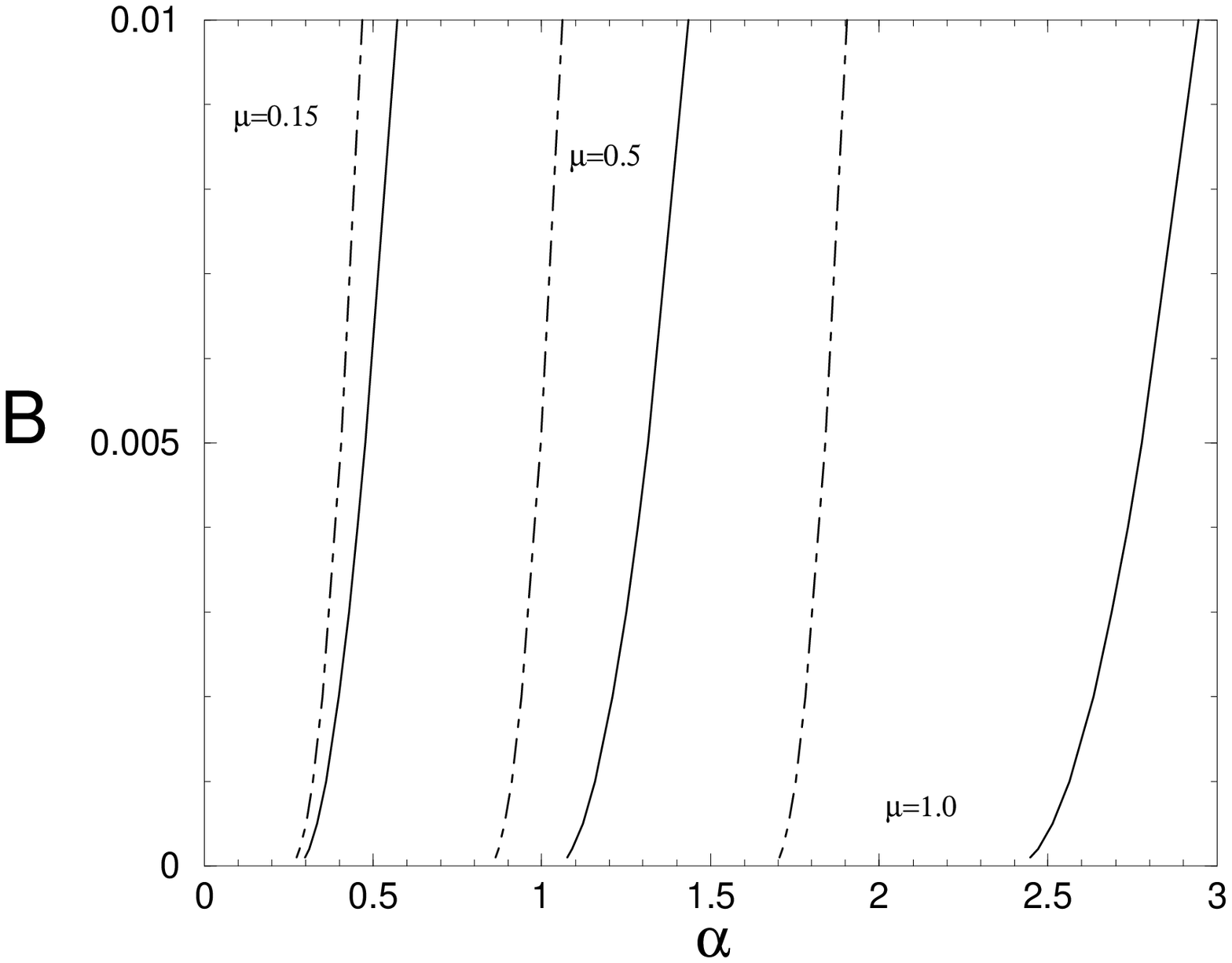}}}} 
\vspace{-0.1cm} 
\caption{Zero binding energy limit of LFD and BS equations (solid line)  
compared with the corresponding non relativistic solution (dot-dashed), 
for $\mu=1$ (a) and for different values of  
$\mu$ (b).}\label{BS_NR_alpha_B}
\end{center} 
\end{figure} 
The solutions for this model in LFD have been obtained in \cite{MC_00}
and turn to be very close to those provided by Bethe-Salpeter equation
in a wide range of coupling constant, well beyond the perturbative region.
We found in particular (see figure \ref{BS_NR_alpha_B}), that for $\mu\ne0$ and
in the zero binding limit, the results for both equations -- indistinguished 
in Figure (b) -- dramatically depart from the  
non-relativistic theory, that is Schr\"odinger equation with a Yukawa potential.
This difference vanishes for $\mu=0$ and increases with $\mu$.
We conclude from that to the irrelevance of a   
non-relativistic approach to describe systems interacting via  
massive fields, what is the case of all the strong interaction physics  
when not described by gluons. 
 
Equations (\ref{S1}) determine the total angular  
momentum $J$ and its projection $\lambda$. They contain interaction and  
this interaction dependence is not easy to deal with. To circumvent this 
difficulty, we make use of the so-called angular condition,  
derived from the transformation properties of the state vector under  
rotations of the light-front plane. In the same way the time dependence is  
given by Schr\"odinger equation, the dependence on the 
the light-front direction $\omega$ is given by \cite{VAK_82}
\vspace{-0.3cm} 
\begin{eqnarray}\label{ang_cond} 
\hat{L}_{\mu\nu}(\omega) \Psi_{\omega}(p)= 
\hat{J}_{\mu\nu}^{int} \Psi_{\omega}(p) 
\end{eqnarray} 
where $\hat{J}_{\mu\nu}^{int}$ is the interacting part of the angular  
momentum tensor $\hat{J}_{\mu\nu}$ and $\hat{L}_{\mu\nu}$ reads 
\vspace{-0.3cm} 
\begin{eqnarray*} 
\hat{L}_{\mu\nu}(\omega)=i \left( \omega_{\mu}  
{\partial \over \partial\omega^{\nu}} -\omega_{\nu}  
{\partial \over \partial\omega^{\mu}}\right)  
\end{eqnarray*} 
Assuming that the state vector satisfies (\ref{ang_cond}), we can  
replace in the Pauli-Lubansky operator $\hat{J}_{\mu\nu}^{int}$ by  
$\hat{L}^{\mu\nu}$. Namely, instead of the dynamical angular  
momentum tensor $\hat{J}^{\mu\nu}$ we introduce the operator $  
\hat{M}_{\mu\nu}= \hat{J}^{0}_{\mu\nu}+\hat{L}_{\mu\nu} $.   
The set $\hat{P}_{\mu}$ and $\hat{M}_{\mu\nu}$ form an algebra of the  
Poincar\'e group in the same way as $\hat{P}_{\mu}$ and  
$\hat{J}_{\mu\nu}$ do. We thus construct a new Pauli-Lubansky vector: $  
\hat{W}_{\mu}={1\over 2} \varepsilon_{\mu\nu\rho\sigma} \hat{P}^{\nu}  
\hat{M}^{\rho\sigma} $ and, provided $\Psi$ verifies the angular  
condition, equations (\ref{S1}) can be replaced by 
\vspace{-0.0cm} 
\begin{equation}\label{W1} 
\hat{W}^2 \Psi = -M^2 J(J+1) \Psi, \quad  
\hat{W}^3 \Psi = M\lambda \Psi  
\end{equation} 
We deal now with an extended Fock space, where $\omega$ is not a fixed  
parameter but a four-vector variable.  
Since $\Psi$ is an eigenvector of $P^{\mu}$, equations (\ref{W1})  
determining the angular momentum are {\it {purely  
kinematical}}. Therefore, their solutions can be found using the  
standard methods of the theory of angular momentum.  
\par  
Now, to construct $J\ne0$ states, we start solving the mass and four-momentum  
equations (\ref{P1}) together with equations (\ref{W1}).  
This procedure gives $2J+1$ solutions which, in a full Fock space, are
degenerated in mass \cite{CDMK_98}. Indeed, it is readily verified 
that the operator  
$\hat{A}^2=(\omega\cdot\hat{W})^2$ commutes with $P^{\mu},\, W^{2},  
\,W_{3}$ and with parity. Therefore the state vector  
satisfying (\ref{P1}) and (\ref{W1}) is  
characterized not only by its mass, momentum and angular momentum, but  
also by the eigenvalues $a^2$ of $\hat{A}^2$: $\hat{A}^2 \Psi_a = a^2  
\Psi_a$. The operator $\hat{A}^2$ has $J+1$ eigenvalues and $2J+1$  
eigenfunctions which are split in two families of opposite parities.   
\par  
$\hat{A}^2$ commutes also with $\hat{S}^2$ but not with the operator  
$\hat{L}_{\mu\nu}-\hat{J}^{int}_{\mu\nu}$, that is the states $\Psi_a$  
do not satisfy the angular condition (\ref{ang_cond}).  The physical  
solution, i.e. the solution satisfying the angular condition, is  
constructed from a superposition of the degenerated states $\Psi_a$: 
\begin{equation}\label{superpos} 
\Psi_{J\lambda}=\sum c_a \Psi_a 
\end{equation} 
\vspace{-0.1cm} 
with coefficients $c_a$ chosen such that $\Psi_{J\lambda}$ satisfies  
(\ref{ang_cond}).  
\par  
Let us summarize the way we construct states with definite angular  
momentum. Instead of dynamical equations (\ref{P1}) and (\ref{S1}), we  
solve equations (\ref{P1}) together with the  
kinematical ones (\ref{W1}). To have the full equivalence between both  
systems of equations, the solutions must be constructed as a  
superposition satisfying the angular condition (\ref{ang_cond}).  
\par  
We propose a simple method to find the right linear combination  
without explicitly solving (\ref{ang_cond}). It is based on the fact 
that, when $k\to 0$, the  
interaction part $J^{int}_{\mu\nu}$ is irrelevant and the angular  
condition reads: $\hat{L}_{\mu\nu} \Psi=0 $.  Thus in this limit, $\Psi$ 
does not depend on the light-front plane direction and hence should not depend  
on $\hat{n}$ anymore. This unambiguously determines the coefficients of  
the superposition. 
 
Our method is hereafter explicited for the case $J=1^{-}$ but the results  
will be given for $J=2$ states as well. We are looking for solution of  
(\ref{P1}) and (\ref{W1}) with $J=1^{-}$, which are also eigenvectors  
of $\hat{A}^2$. The general form of these solutions reads  
$\vec{\psi}_{a}(\hat{k},\hat{n})=  
\vec{\chi}_{a}(\hat{k},\hat{n})\,g_a(k,z) $ where $z=\hat{n}\cdot  
\hat{k}$ and $\vec{\chi}_{a}(\hat{k},\hat{n})$ are given by
\vspace{-0.2cm} 
\begin{eqnarray*}  
\vec{\chi}_{0}(\hat{k},\hat{n}) = 3z\hat{n} \qquad \qquad 
\vec{\chi}_{1}(\hat{k},\hat{n}) = \frac{3\sqrt{2}}{2}  
(\hat{k}-z\hat{n})
\end{eqnarray*}  
The scalar functions $g_0(k,z)$ and $g_1(k,z)$
satisfy: 
\vspace{-0.1cm} 
\begin{eqnarray}  
&&[4(\vec{k}^2 +m^2)-M_0^2]zg_0(k,z)=-\frac{m^2}{2\pi^3}  
\int\frac{d^3k'}{\varepsilon_{k'}} 
V (\vec{k}\,',\vec{k},\hat{n},M_0) z'g_0(k',z') 
\label{g0}\\ 
&&[4(\vec{k}^2 +m^2)-M_1^2](1-z^2)\,g_1(k,z) 
\nonumber\\ 
&& 
\qquad \qquad \qquad \quad \, 
=-\frac{m^2}{2\pi^3} \int\frac{d^3k'}{\varepsilon_{k'}}  
V(\vec{k}\,',\vec{k},\hat{n},M_1) 
(\hat{k}\cdot\hat{k}'-zz')g_1(k',z') 
\label{g1}  
\end{eqnarray}  
As a consequence of the Fock space truncation, the mass degeneracy of  
states with different $a$ is violated. These states are split:   
different $a$'s correspond to different masses. Therefore equations  
(\ref{g0}) and (\ref{g1}) give two solutions with different masses  
$M_0$ and $M_1$. It was found in \cite{CMP_00} that the splitting reduces 
when the two-boson exchanges, incorporating effectively higher Fock states, 
are added to the kernel. 
According to (\ref{superpos}), the physical solution is  
\begin{equation}\label{phys} 
\vec{\psi}(\vec{k},\hat{n})=c_0\vec{\psi}_0(\vec{k},\hat{n}) 
+c_1\vec{\psi}_1(\vec{k},\hat{n}) 
\end{equation} 
\vspace{-0.1cm} 
with $c_0$ and $c_1$ determined by the condition that in the vicinity  
of $k=0$ $\vec{\psi}(\vec{k},\hat{n})\propto k$ is independent of  
$\hat{n}$, which leads to 
\[ c_0= h_1/\sqrt{2+h_1^2} \qquad \qquad c_1=  
\sqrt{2}/\sqrt{2+h_1^2} \]
with $h_1=\lim_{k \to 0}{g_1(k,z)\over  
g_0(k,z)}$.  The coefficients $c_0$ and $c_1$, which in principle
depend on $\alpha$, were found to be very  
close to $\sqrt{\frac{1}{3}}$ and $\sqrt{\frac{2}{3}}$ (with the  
accuracy $\approx 1\%$).  We emphasize that we take the superposition  
(\ref{phys}) inspite of the fact that $\vec{\psi}_{0,1}$ correspond to  
different masses and the mass of the physical solution is 
\[ M^2=c_0^2M_0^2 +c_1^2M_1^2\]  
 
\begin{figure}[hbtp]  
\begin{minipage}{60mm} 
{\mbox{\epsfxsize=6.8cm\epsfysize=8cm\epsffile{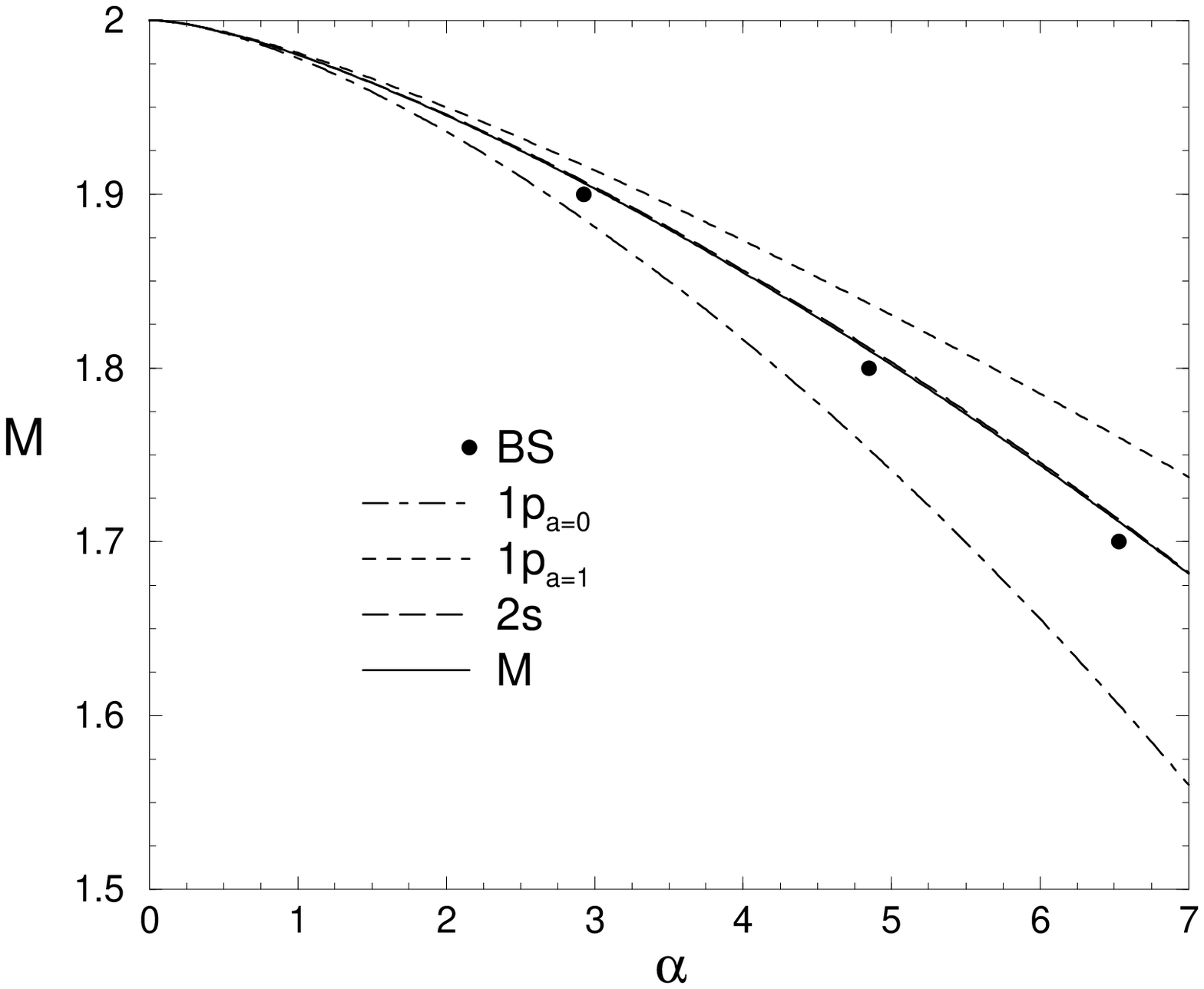}}} 
\caption{$2s$ and $1p$ degeneracy ($\mu$=0)}\label{2s1p_Mphys}
\end{minipage} 
\hspace{0.9cm} 
\begin{minipage}{60mm} 
{\mbox{\epsfxsize=6.8cm\epsfysize=8cm\epsffile{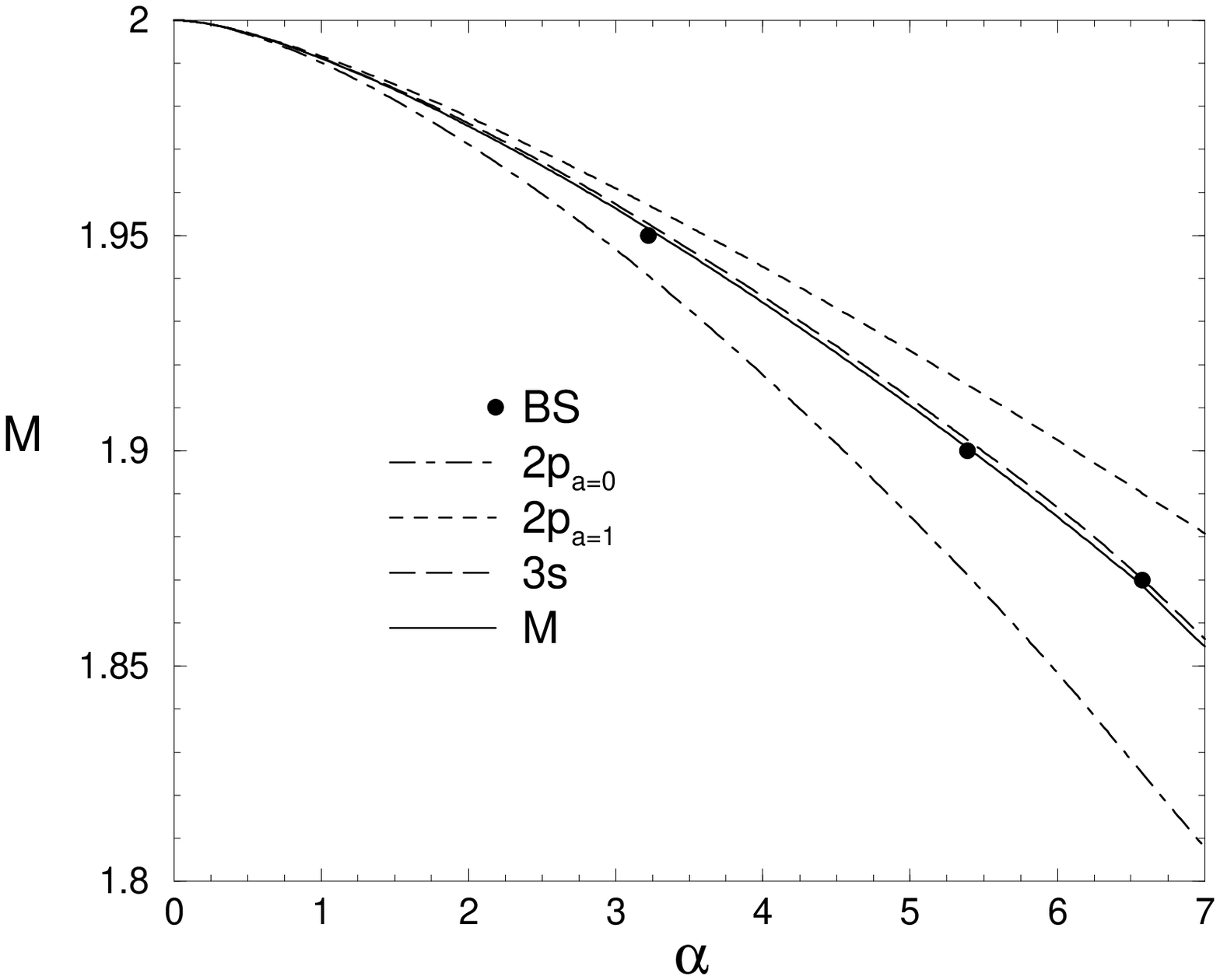}}} 
\caption{$3s$ and $2p$ degeneracy ($\mu$=0)}\label{3s2p_Mphys} 
\end{minipage} 
\end{figure} 

In Figure \ref{2s1p_Mphys} the masses of $2s$ state are shown, together  
with $M_0$, $M_1$ and the physical mass of the $1p$ state $M$, as  
functions of the coupling constant $\alpha=g^2/16\pi m^2$. It is seen  
that the degeneracy which exists between $2s$ and $1p$ states in the  
case $\mu=0$ is restored up to a high accuracy. This is also the case  
for $3s$ and $2p$  states in Figure \ref{3s2p_Mphys}. 
Furthermore the physical masses $M$ are shown to be close to those  
provided by the resolution of Bethe-Salpeter equation, whereas the two  
approaches do not have the same diagrammatical content.

\begin{figure}[hbtp] 
\begin{center} 
{\mbox{\epsfxsize=7.5cm\epsfysize=8cm 
\epsffile{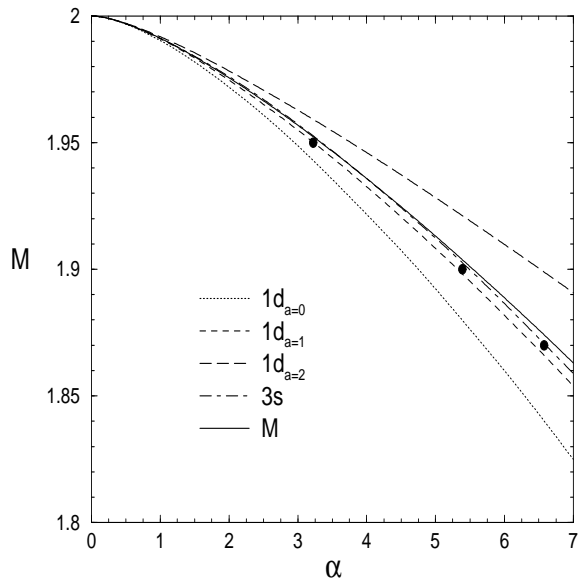}}} 
\end{center} 
\caption{Restoration of $3s$ and $1d$ degeneracy for $\mu=0$}\label{3s1d_Mphys} 
\end{figure}

We display in Figure \ref{3s1d_Mphys} the equivalent results for $J=2$, i.e. 
the $3s$-$1d$ degeneracy. Let us however mention that the case $J>1$ has
some specificities which require some care, and will be presented elsewhere.  

\bigskip 
 
As a summary, by studying the zero binding limit, we come to the conclusion that  
composite systems strongly interacting via a massive exchange cannot be properly  
described by a non relativistic dynamics \cite{MC_00}. Furthermore, we have developed
a method to construct non zero angular momentum states in the  
explicitly covariant LFD framework. The results presented for $J=1$  
and $J=2$ show that the rotational degeneracy in the case of a massless  
exchange is restored up to a high level of accuracy, and that binding  
energies are close to Bethe-Salpeter ones.  
The construction of $J\ne0$ states proposed has been applied 
to a scalar model: the next challenge is the description of 
fermionic systems.


\begin{thebibliography}{100} 
\bibitem{CDMK_98} J. Carbonell et al., Phys. Rep. {\bf300} (1998) 215    
\bibitem{BPP_98} S.J. Brodsky, H-C. Pauli and S. Pinsky, Phys. Rep.  
{\bf301} (1998) 299    
\bibitem{MC_00} M. Mangin-Brinet and J. Carbonell,  
Phys. Lett. {\bf B474} (2000) 237 
\bibitem{VAK_82}  
V.A. Karmanov, Sov. Phys. JETP 56 (1982) 1  
\bibitem{CMP_00} J.R. Cooke, G.A. Miller and D. Phillips,  
Phys. Rev. {\bf C61} (2000) 064005 
\end{thebibliography}
\end{document}